\documentclass[10pt,letterpaper]{article}
\usepackage{upgreek}
\usepackage[numbers,square,comma,sort&compress]{natbib}
\usepackage{opex3}
\begin{document}

\title{Tracking Rotational Diffusion of Colloidal Clusters
}

\author{Gary L. Hunter$^1$*, Kazem V. Edmond$^1$, Mark T. Elsesser$^2$, and Eric R. Weeks$^1$}

\address{$^1$ Department of Physics, Emory University, Atlanta, GA 30322, USA}
\address{$^2$ Center for Soft Matter Research, Department of Physics, New York University, New York, NY 10003, USA}
\email{*glhunter@gmail.com} 

\begin{abstract}
We describe a novel method of tracking the rotational motion of clusters of colloidal particles.  Our method utilizes rigid body transformations to determine the rotations of a cluster and extends conventional proven particle tracking techniques in a simple way, thus facilitating the study of rotational dynamics in systems containing or composed of colloidal clusters.  We test our method by measuring dynamical properties of simulated Brownian clusters under conditions relevant to microscopy experiments.  We then use the technique to track and describe the motions of a real colloidal cluster imaged with confocal microscopy.  
\end{abstract}
%

\ocis{(000.2170) Equipment and techniques; (000.4430) Numerical approximation and analysis; (100.2000) Digital image processing; (100.4999) Pattern recognition, target tracking;  (180.1790) Confocal microscopy; (180.6900)  Three-dimensional microscopy. } 



\section{Introduction}

Suspensions of spherical colloidal particles have proven a valuable model system for understanding  many complex phenomena. Perhaps most notably, the model has provided insight into dynamical processes within different phases of matter \cite{kose1973jcis,ackerson1986,murray1987,chaikin1989,Pusey1986,pusey1987,pusey1994jcmp,weeks2000sci}, as well as various mechanisms involved during phase transitions \cite{savagesci2006,vanwinkle1986, murray1990,grier94,gasser01,hernandez09}.
A combination of digital video microscopy and computerized particle tracking algorithms \cite{crocker1996jcis,weeksmicroscopy1} has allowed for the direct visualization of such colloidal systems and measurement of static and dynamical properties under many experimental conditions~\cite{ grier00prl,nugent2007prl, besseling07, schall2007sci,chen10}. Given the spherical symmetry of the particles, most previous studies have understandably focused only on understanding translational dynamics, though there have been a few studies on rotational dynamics in dilute systems of spheres \cite{martin2006rot,anthonymoon2006}, and on the translational and rotational dynamics of anisotropic colloidal particles \cite{yunker2011,granick2006rot,lubensky2006sci, solomon2007rods}.

\begin{figure}[htbp]
\begin{center}
\includegraphics[width=.95\textwidth]{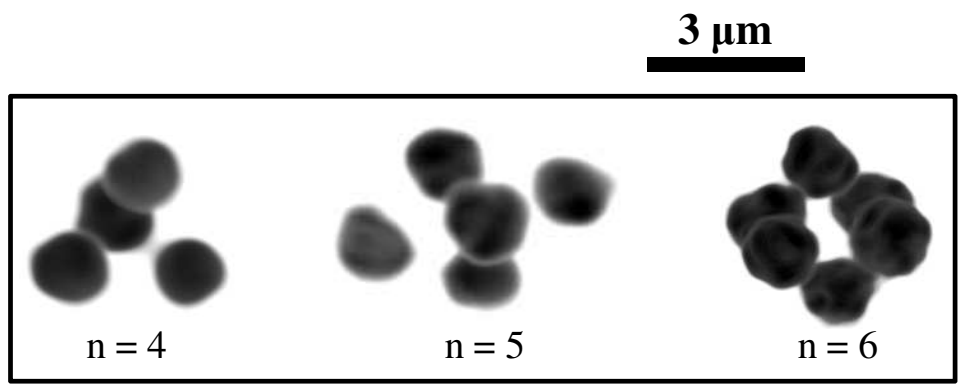}
 \caption{Volumetric images of colloidal clusters with $n=4,~5,$ and $6$ from 3D confocal micrographs.   Images have been filtered and enhanced to allow easy visualization of the 3-dimensional structures.  The cases $n=4$ and $n=5$ are accurate representations of the simulated tetrahedra and pentahedra discussed in the text.  Individual particles are approximately 2~$\mu$m in diameter.}
\end{center}
\label{fig:clusters}
\end{figure}

Within the past decade, researchers have developed a variety of techniques for synthesizing clusters of colloidal particles with a wide range of reproducible morphologies \cite{vinnysci03,elsesserlang2011}, several of which are shown in Fig. \ref{fig:clusters}.  Colloidal clusters have the potential to extend the colloidal model past one of simple spherical particles and into a realm where the collective dynamics of particles with complex shapes, more representative of molecules, can be studied \cite{vanblaasci03, vanblaanat06, glotzer2007natmat}. 

Conventional particle tracking methods \cite{crocker1996jcis,weeksmicroscopy1} are designed to follow the translational motions of individual particles, and so are immediately applicable to colloidal clusters, provided the particles can be reliably distinguished.  However, a description of dynamics within systems composed of or including clusters is incomplete without knowledge of how rotational degrees of freedom are explored.  Such an understanding could provide further insight into fundamental behaviors of systems with orientational order, such as liquid crystals, or systems subjected to external fields and anisotropic flows \cite{ solomon2007rods}. To our knowledge, there has been only one study that combined video microscopy with particle tracking to measure the rotational motion of colloidal clusters~\cite{anthony2008decoupling}, using different methods than described in this work.  That study focused on dilute systems of planar clusters undergoing two-dimensional diffusion near a boundary, and revealed a decoupling of translational and rotational diffusion due to hydrodynamic effects.

We present here a simple and generalized method to track and calculate the two- or three-dimensional rotational motions of clusters of colloidal particles.  Our method uses existing particle tracking routines and rigid body transformations to measure the changes in orientation of a cluster over time.  To demonstrate the effectiveness and accuracy of the method, we simulate the motion of colloidal clusters under conditions that are relevant to common microscopy experiments.  We then use our method to measure the rotational diffusion coefficient of a real colloidal cluster, and compare the results of both simulations and experiments to other methods of measuring rotational diffusion.

\section{Calculating Rotations}
\label{sec:calcrots}
Our procedure for calculating rotational displacements of clusters is based on a method by Challis for determining rigid body transformations between reference frames \cite{challisrot}.  Challis' procedure was originally intended for comparing osteometric measurements in biomechanical analyses.  As we will show, the method can be intuitively adapted to measure rotational dynamics.  First, we give a brief reprise of Challis' method, and then describe how it is used to study the systems mentioned here.

\subsection{Challis' Procedure for Coordinate Transformations}
\label{sec:challis}
Given a set of points which have coordinates $\{\mathbf{x}_i\}$ measured in one reference frame and coordinates $\{\mathbf{y}_i\}$ measured in a second frame, there exists a transformation

\begin{equation}
\label{eqn:chalreftran}
\mathbf{y}_i = s\mathbf{Rx}_i + \mathbf{v},
\end{equation}
where subscript $i$ refers to the $i$-th point in the set, $s$ is a scale factor, $\mathbf{R}$ is a 3$\times$3 rotation matrix, and $\mathbf{v}$ is the vector separation of the two reference frames.  For our purposes, we may set the scale factor to unity and assume that both coordinate frames share a common origin, thereby setting all elements of $\mathbf{v}$ to zero.

We are therefore left with 
\begin{equation}
\label{eqn:chalrtran}
\mathbf{y}_i = \mathbf{Rx}_i,
\end{equation}
and only the rotation matrix to describe the transformation between coordinate frames.  The rotation matrix is an orthonormal matrix with the properties
\begin{equation}
\label{eqn:orthonormal}
\mathbf{R}\mathbf{R}^{-1} = \mathbf{R} \mathbf{R}^\mathrm{T}  = \mathbf{R}^\mathrm{T} \mathbf{R}  =\mathbf{E},
\end{equation}

\begin{equation}
\label{eqn:determinant}
\det{(\mathbf{R})} = +1,
\end{equation}
where $\mathbf{E}$ is the identity matrix and $\det{()} $ denotes the determinant.

For a set of $n$ points, $\mathbf{R}$ can be calculated using a least squares approach.  This method minimizes the quantity
\begin{equation}
\label{eqn:leastsquares}
\displaystyle \frac{1}{n}\sum_{i=1}^{n} \left[ \mathbf{y}_i - \mathbf{R}\mathbf{x}_i \right]^\mathrm{T} \left[ \mathbf{y}_i - \mathbf{R}\mathbf{x}_i\right].
\end{equation}
Ignoring the factor of $1/n$, expansion of Eq. (\ref{eqn:leastsquares}) yields

\begin{equation}
\label{eqn:leastsquares2}
\displaystyle \sum_{i=1}^{n} (\mathbf{y}_i)^\mathrm{T} \mathbf{y}_i + (\mathbf{x}_i)^\mathrm{T} \mathbf{x}_i - 2(\mathbf{y}_i)^{\mathrm{T}}\mathbf{R}\mathbf{x}_i.
\end{equation}
Given that all $\mathbf{x}_i$ and $\mathbf{y}_i$ are fixed, minimizing Eq. (\ref{eqn:leastsquares2}) is therefore equivalent to maximizing
\begin{equation}
\label{eqn:maximize}
\displaystyle \sum_{i=1}^{n}(\mathbf{y}_i)^{\mathrm{T}}\mathbf{R}\mathbf{x}_i=     
\mathrm{Tr}\left(\mathbf{R}^{\mathrm{T}} \frac{1}{n} \sum_{i=1}^{n} \mathbf{y}_i(\mathbf{x}_i)^{\mathrm{T}} \right)=
 \mathrm{Tr}(\mathbf{R}^\mathrm{T} \textbf{C}),
\end{equation}
where $\textbf{C}$ is the cross-dispersion matrix calculated from

\begin{equation}
\label{eqn:crossdispersion}
\textbf{C} = \displaystyle \sum_{i=1}^{n} \mathbf{y}_i (\mathbf{x}_i)^\mathrm{T}.
\end{equation}

At this point, a singular value decomposition is performed on $\textbf{C}$ such that
\begin{equation}
\label{eqn:svd}
\textbf{C} = \textbf{UWV}^\mathrm{T},
\end{equation}
where $\mathbf{W}$ is a diagonal matrix containing the singular values of $\mathbf{C}$, and $\mathbf{U}$ and $\mathbf{V}$ are orthogonal matrices.   As was shown in \cite{challisrot}, upon maximizing Eq. (\ref{eqn:maximize}), $\mathbf{R}$ is given by
\begin{equation}
\label{eqn:rotmatfin}
\mathbf{R} = \textbf{U} \left[ \begin{array}{ccc}
1 & 0 & 0 \\
0 & 1 & 0 \\
0 & 0 & \det(\textbf{U} \textbf{V}^\mathrm{T}) \end{array} \right]  \textbf{V}^\mathrm{T}.
\end{equation}
This procedure is applicable to all non-colinear sets of points with $n \geq 3$.  

\subsection{Application to Colloidal Clusters}
\label{sec:challisonclusters}

Particle tracking \cite{crocker1996jcis,weeksmicroscopy1} yields vector coordinates $\mathbf{x}'_i$ for every particle $i$ over a distinct set of times. Hence, the first step in tracking rotational motion of a cluster is to track the translational motion of each particle within the cluster.  For each cluster, we first determine the center of mass $\mathbf{x}_{CM}$ at a given time and subtract this quantity from the coordinates of particles belonging to the cluster, thereby removing any translational motion.  We are left with new coordinates $\mathbf{x}_i$ in the center of mass frame,
\begin{equation}
\label{eqn:comframe}
\mathbf{x}_i = \mathbf{x}'_i - \mathbf{x}_{CM}.
\end{equation}
This step is equivalent to setting the elements of $\mathbf{v}$ to zero in Eq. (\ref{eqn:chalreftran}).  With translational motion removed, we may apply  Eq. (\ref{eqn:chalrtran}) with a slightly different interpretation.  Rather than representing a transformation between coordinate frames, we may understand $\mathbf{R}$ as describing the rotational trajectory of a particle with inital position $\mathbf{x}_i^0$ to a final position $\mathbf{x}_i$ such that
\begin{equation}
\label{eqn:rotmat}
\mathbf{x}_i = \mathbf{R} \mathbf{x}_i^0.
\end{equation}

Therefore, we may use Challis' procedure to calculate a rotation matrix for each pair of successive times $\left[ t, t + \Delta t \right]$.  With the complete set of rotation matrices, $\{ \mathbf{R}^k \}$, we may reconstruct the entire trajectory of a particle about the cluster center of mass by computing the product of successive rotations.  Given $\mathbf{x}_i^0$, the position of a particle at some later time $t$ can be calculated as
\begin{equation}
\label{eqn:rotprod}
\mathbf{x}_i(t) = \mathbf{R}^{t-\Delta t} \mathbf{R}^{t-2\Delta t} ... \ \mathbf{R}^{0} \mathbf{x}_i^0 = \prod_{k} \mathbf{R}^{k}\mathbf{x}_i^0,
\end{equation}
where the index $k$ enumerates the rotation between successive times. 

The advantage of calculating $\{\mathbf{R^k}\}$ is that it describes the collective behaviors of particles within a cluster, rather than a property of any individual particle.  For example, knowledge of $\{\mathbf{R^k}\}$ for a cluster allows for immediate calculation of the motions of any particular particle about the center of mass, or the motion of the cluster about any arbitrary axis of rotation.  Diffusive anisotropic clusters with large aspect ratios rotate more freely about a long axis than about a short axis.  Given $\{ \mathbf{R^k} \}$, however, one needs only the initial orientation of these axes to compute and compare the motions around them.
 

\section{Tests of the Prescribed Method}\label{sec:tests}

For simplicity and to better reproduce the data collection process in typical microscopy experiments, we adopt a length scale of microns and a time scale measured in timesteps (ts), which is equivalent in microscopy to video frames or image stacks.  

In conventional particle tracking experiments, it is important to minimize and understand the uncertainty, i.e. the noise, inherent in locating a particle.  Typical microscopy experiments combine high magnification optics with CCD cameras to record raw digital images. In the absence of other sources of noise, the uncertainty in particle position, i.e. the minimum noise level, depends on the optical resolution [pixels/distance] of the instrumentation and the size [pixels] of the object being tracked.  Standard image processing and particle tracking techniques can locate the centers of particles to within $\approx 1/N$ of a pixel, where $N$ is the width of the object in pixels.  Optical resolution varies between experimental set-ups but is typically in the range of $0.2 \ \mu \mathrm{m} / \mathrm{pixel} $.  The minimum uncertainty in particle position is the product of these factors.  For example, observing a $10$ pixel wide object with an optical resolution of $0.2 \ \mu \mathrm{m} / \mathrm{pixel} $ leads to a lower limit of $\approx 20~$nm uncertainty  in particle position.  Other sources of noise, such as stray light entering the microscope, noise within the CCD camera itself, etc., slightly increase the uncertainty in particle position and further limit particle tracking resolution. 

To test our method, we simulate the rotational Brownian motion of tetra- and pentahedral clusters with different rotational diffusion coefficients, $D_R$, and different levels of noise, $\sigma_x$.  We first generate noise-free cluster trajectories. For tetrehedra, we place particles at initial coordinates

\begin{equation}
\label{eqn:tetracoords}
 \begin{array}{c}
(R/\sqrt{3})\cdot(1, \pm 1, \pm 1) ,\\
(R/\sqrt{3})\cdot (-1, \pm 1, \mp 1) ,\\
\end{array}
\end{equation}
where $R$ is the distance from a particle center to the cluster center of mass.  For pentahedra, we use initial coordinates

\begin{equation}
\label{eqn:pentacoords}
 \begin{array}{c}
R\cdot(1, 0,0) ,\\
(R/2)\cdot (-1, \pm \sqrt{3}, 0) ,\\
R\cdot(0,0, \pm 1) .
\end{array}
\end{equation}
Once initialized, we evolve each simulation for $10^{4}$ time steps.

At each time step, we select three random angles, $\alpha, \beta, \gamma$,  from a Gaussian distribution with a standard deviation of $\sqrt{2 D_R}$.  This distribution ensures that the simulated dynamics will be in agreement with the Stokes-Einstein-Debye relation discussed later [Eq. (\ref{eqn:sed})].  Each particle in a tetrahedron is rotated by an angle $\gamma$ about the $z$-axis, then by an angle $\beta$ about the $y$-axis, and finally by angle $\alpha$ about the $x$-axis to produce the tetrahedron at the subsequent time.  The rotation matrices used are 

\begin{eqnarray*}
R_x(\alpha) = \left( \begin{array}{ccc} 1 & 0 & 0 \\ 0 & \cos{\alpha} & -\sin{\alpha}  \\ 0  & \sin{\alpha} & \cos{\alpha} \end{array} \right),
\ \ \ R_y(\beta) = \left( \begin{array}{ccc} \cos{\beta} & 0 & \sin{\beta} \\ 0 & 1 & 0  \\ -\sin{\beta}  & 0 & \cos{\beta} \end{array} \right),
\end{eqnarray*}

\begin{eqnarray}
R_z(\gamma) = \left( \begin{array}{ccc} \cos{\gamma} & -\sin{\gamma} & 0 \\ \sin{\gamma} & \cos{\gamma} & 0  \\ 0  & 0 & 1 \end{array} \right).
\end{eqnarray}
Thus, given an initial position vector $\mathbf{x}^0$, the subsequent position vector is $\mathbf{x} = \mathbf{R}_x \mathbf{R}_y \mathbf{R}_z \mathbf{x}^0$.  

Trajectories generated in this way are noise-free in the sense that they are absent of uncertainty in particle position to within machine precision. To mimic the type of experimental noise previously mentioned, we post-process the trajectories by adding Gaussian random numbers, with standard deviation $\sigma_x$, to the particle coordinates.  The levels of noise presented here correspond to uncertainties of $\sigma_x \in \{10,~30,~50,~70,~100 \}$ nm in the $x$-, $y$-, and $z$-directions.  Experimental uncertainties are typically within the range of 20-60 nm, and so the levels studied here are relevant to microscopy experiments.  After the noise is added, we apply our method of measuring rotational motion in order to gauge the effect of experimental noise on results.


\section{Analysis of Rotational Motion}
\label{sec:analysis}

In this section, we calculate the rotational motions of simulated colloidal clusters using the matrix methods described above.  We focus on a method that uses the rotation matrices to determine the motion of a fictional orientation vector attached to the cluster.  The rotational displacements of one or more such vectors about the center of mass can be used to compare and contrast motions about different axes, which is especially useful when the objects studied are anisotropic and motions about axes are expected to differ \cite{mazza2006,mazzapre2007}.  

We note that a second method exists to calculate rotational displacements using solely the rotation matrix.  This method determines the rotational axis from $\mathbf{R}$ and can then be used to calculate the magnitude of an angular displacement.  While a mathematically direct and general approach, it describes only the average cluster dynamics and can provide no insight into how motions about different axes vary.  Given these limitations,  we relegate a description of this method to the appendix.

%
%


\subsection{Rotations via Orientations}
\label{sec:rotviaorient}
To analyze the rotations of a cluster, we consider the motions of an initial orientational unit vector  $\hat{\mathbf{p}}^0$ fixed to the cluster.  We determine its orientation at a later time $t$ by applying the set of rotations such that
\begin{equation}
\label{eqn:rphat}
\hat{\mathbf{p}}(t) = \prod_{k} \mathbf{R}^{k}\hat{\mathbf{p}}^0,
\end{equation}
in a similar way as in Eq. (\ref{eqn:rotprod}).   Although there are no constraints on what one may select $\hat{\mathbf{p}}^0$ to be, some choices may be more enlightening than others.  For example, the cluster $n=5$ shown in Fig. \ref{fig:clusters} has distinct long and short axes, and so one expects slower diffusion about the short axis.  To quantify how rotational dynamics about these axes differ, one could choose two $\hat{\mathbf{p}}^0$ to study separately:  one choice of $\hat{\mathbf{p}}^0$ perpendicular to the long axis; a second perpendicular to the short.  Such a procedure would yield information relating to motions about the long and short axes, respectively.

In line with~\cite{mazza2006,mazzapre2007,kobpre1997}, we define a vector rotational displacement

\begin{equation}
\label{eqn:vecdtheta}
\displaystyle \mathbf{\vec{\upvarphi}}(t) = \int_{0}^{t}\vec{\omega}(t')dt'
\end{equation}

\noindent in the time interval $\left[ 0, t\right]$.  The vector $\vec{\omega}(t')dt'$ has a direction given by $\hat{\mathbf{p}}(t') \times \hat{\mathbf{p}}(t'+dt')$ and  magnitude $|\vec{\omega}(t')dt' |  = \cos^{-1} \left[ \hat{\mathbf{p}}(t') \cdot \hat{\mathbf{p}}(t' + dt') \right] $, which is the angle subtended by $\hat{\mathbf{p}}$ during this time interval. 

To illustrate the meaning of $\mathbf{\vec{\upvarphi}}(t)$, consider an object with constant angular velocity $\vec{\mathbf{\omega}} = \omega \hat\mathbf{z}$. Over a time $\Delta t$, the rotational vector displacement is given by $\vec{\upvarphi}(\Delta t) = \vec{\omega} \Delta t= (0,0,\omega \Delta t)$. Therefore in general, $\mathbf{\vec{\upvarphi}}(t)$ has components in each of the Cartesian axes, $(\upvarphi_x,\upvarphi_y,\upvarphi_z)$, corresponding to cumulative rotations about those axes.

\begin{figure}
\begin{center}\includegraphics[width=.99\textwidth]{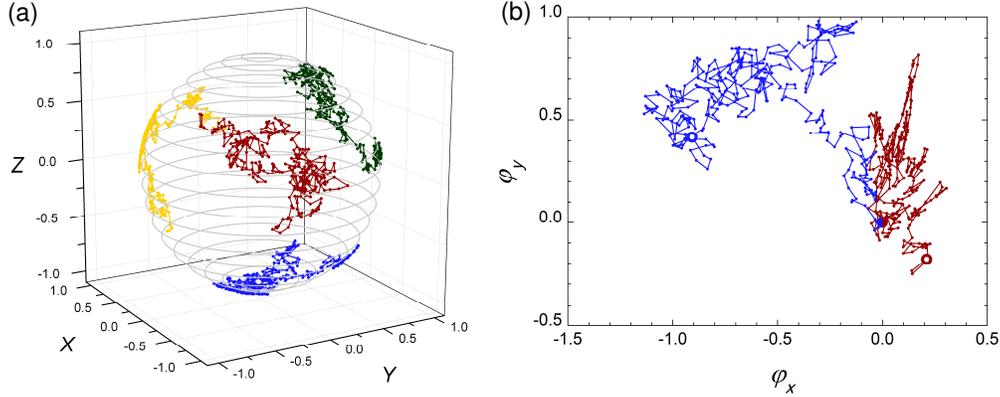}
\end{center}
\caption{(a) Trajectories of particles within a simulated Brownian tetrahedral cluster projected onto the surface of a unit sphere. Colors represent different particles, i.e. different choices for $\hat{\mathbf{p}}^0$.  (b) 2d projection of two trajectories through rotation space $(\upvarphi_x,\upvarphi_y)$. Colors correspond to the same particles in (a).  Both trajectories in (b) begin at $(0,0)$ and end at open circles.}
\label{fig:orient_traj}
\end{figure}

In Fig. \ref{fig:orient_traj}(a), we show the orientational trajectories of particles within a simulated tetrahedron projected onto the surface of a unit sphere. Fig. \ref{fig:orient_traj}(b) shows 2D projections of trajectories for two $\hat{\mathbf{p}}^0$ through the rotation space described above. As shown in Fig. \ref{fig:orient_traj}, even though the cluster is a solid body, trajectories of individual particles differ due to rotations of the cluster about random axes.

Given the definition of $\vec{\upvarphi}(t)$ in Eq. (\ref{eqn:vecdtheta}), we may define an unbounded mean square angular displacement (MSAD), akin to a translational mean square displacement, as
\begin{equation}
\label{eqn:msad}
\langle \Delta \vec{\upvarphi}^2 (\Delta t) \rangle = \langle [\mathbf{\vec{\upvarphi}} (t+\Delta t) - \mathbf{\vec{\upvarphi}} (t)]^2 \rangle,
\end{equation}
where the angle brackets indicate an average over all equivalent lag times $\Delta t$.  In three dimensions the Stokes-Einstein-Debye relation states that the MSAD grows as
\begin{equation}
\label{eqn:sed}
\langle \Delta \vec{\upvarphi}^2 (\Delta t) \rangle = 4D_R \Delta t,
\end{equation}
\noindent where $ D_R$ is the rotational diffusion coefficient.  In this paper, we focus on measuring the MSADs of clusters, but we point out that other techniques exist to quantify rotational dynamics.  For example, by observing the decay of an orientational correlation function $\langle \hat{\mathbf{p}}(t+\Delta t) \cdot  \hat{\mathbf{p}}(t) \rangle $ one can measure $D_R$~\cite{berne68,williams78}.  Our method can be applied, in this case, to compute $\hat{\mathbf{p}}(t)$ as in Eq. (\ref{eqn:rphat}).

\begin{figure}
\begin{center}\includegraphics[width=.9\textwidth]{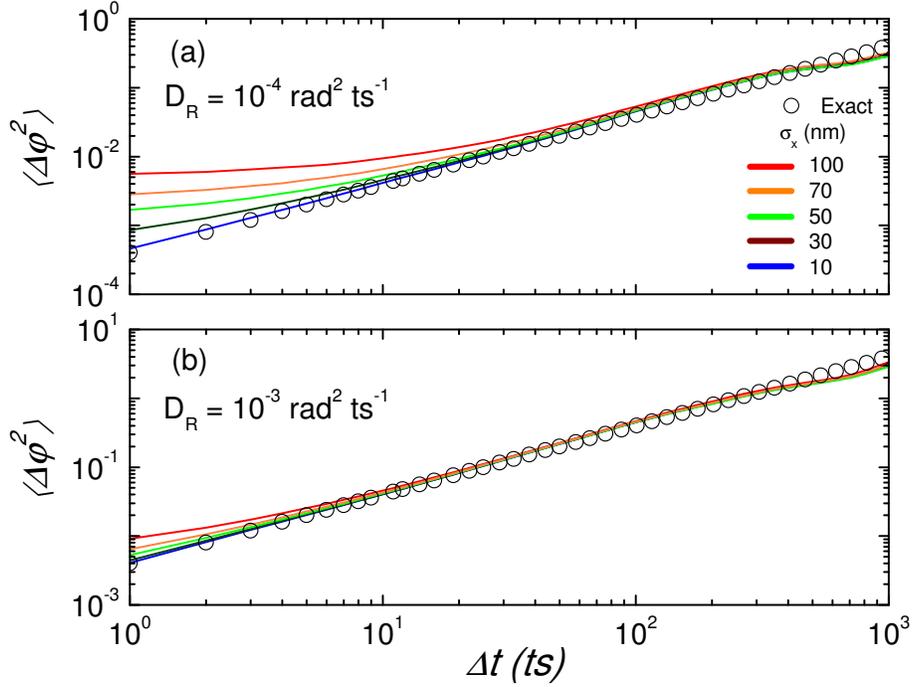}
\end{center}
\caption{Mean square angular displacements of simulated tetrahedral clusters for different noise levels $\sigma_x$ and diffusion coefficients of (a) $10^{-4} ~\mathrm{rad}^2 / \mathrm{ts} $ and (b) $10^{-3} ~\mathrm{rad}^2/ \mathrm{ts}$.  Open circles are the theoretical MSAD based on Eq. (\ref{eqn:sed}).  Deviations from linearity at small $\Delta t$ demonstrate the effect of noise when resolving small rotations.  Deviations at large $\Delta t$, however, are the result of low statistics at these lag times. }
\label{fig:msadnoise}
\end{figure}

In Fig. \ref{fig:msadnoise}(a) \& (b), we show the MSAD of two simulated tetrahedral clusters with diffusion constants $10^{-4} ~\mathrm{rad}^2 /  \mathrm{ts}$  and $10^{-3}~\mathrm{rad}^2 /  \mathrm{ts}$ respectively and different levels of noise.  The influence of noise is apparent in Fig. \ref{fig:msadnoise} as deviations from linearity at small $\Delta t$.  Eventually, the MSAD recovers the true diffusive behavior because the cluster has made rotations large enough to distinguish from the noise.

With no experimental noise, a log-log plot of MSAD as a function of $\Delta t$ will be a straight line with a slope of unity, as indicated by the open circles in Figs.~\ref{fig:msadnoise}(a) \& (b). However, in the presence of noise in particle positions, rotations cannot be accurately resolved below a certain threshold, $\Phi$.  For example, a stationary cluster will appear to make small, but fictional, rotations as a result of this noise, and measurements of the MSAD will yield
\begin{equation}
\langle \Delta \vec{\upvarphi}^2 (\Delta t) \rangle = \Phi^2,
\label{eqn:msadnoise}
\end{equation}

\noindent where  $\Phi^2$ is independent of $\Delta t$ due of the lack of correlations in noise.

 In the case of translational diffusion, one expects that the noise in each direction will contribute an error of $2\sigma_x^2$ to the mean-square displacement \cite{crocker1996jcis}.  In terms of an angular uncertainty, this contribution is diminished by a factor of $R^2$, where $R$ is the average distance from a particle to the cluster center of mass.  Our matrix method reduces this uncertainty further by a factor of $n$.  Thus, when only static noise is present, one expects for a diffusing cluster
\begin{equation}
\label{eqn:expmsad}
\langle \Delta \vec{\upvarphi}^2 (\Delta t) \rangle = 4D_R \Delta t +\Phi^2,
\end{equation}
where
\begin{equation}
\label{eqn:chidefined}
\Phi^2 = 6\sigma_x^2/nR^2.
\end{equation}

\begin{figure}
\begin{center}\includegraphics[width=.99\textwidth]{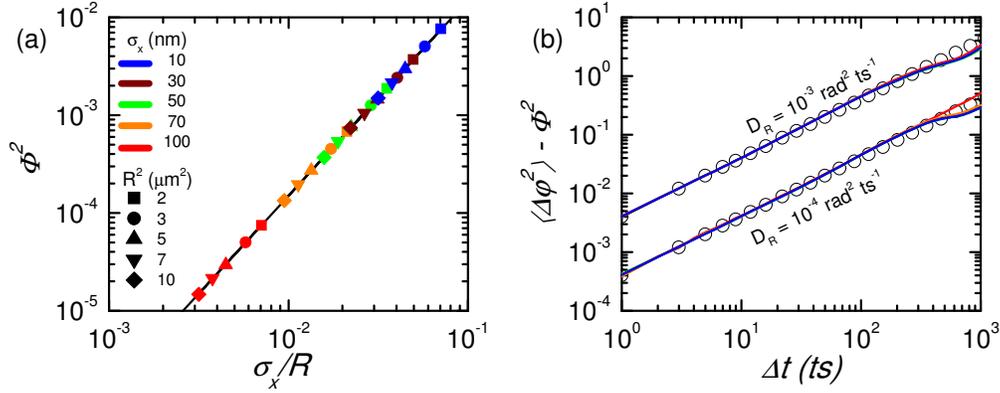}
\end{center}
\caption{(a) Measured values of $\Phi^2$ for non-diffusing tetrahedral clusters.  Colors indicate noise levels and symbols indicate cluster size.  Solid line is the prediction $\Phi^2 = 6 \sigma_x^2 / nR^2$. (b) Same data in Fig. \ref{fig:msadnoise} with the appropriate $\Phi^2$ for each noise level subtracted. $\Phi^2$ for pentahedral clusters collapses in the same way.}
\label{fig:chi2}
\end{figure}

To test this assertion, we add noise to simulations of stationary ($D_R = 0$) tetrahedral clusters and calculate the MSADs.  As in Eq. (\ref{eqn:msadnoise}), the MSADs are constant over time.  We take $\Phi^2$ to be the value of the MSAD as $\Delta t \rightarrow 0$. Shown in Fig. \ref{fig:chi2}(a), Eq.~(\ref{eqn:chidefined}) accurately describes the static angular uncertainty for a wide range of tetrahedral cluster sizes, $R$, and noise levels.  Fig. \ref{fig:chi2}(b) shows all of the previously measured MSADs in Fig.~\ref{fig:msadnoise} plotted with the noise subtracted.  This precisely collapses the MSADs to the true values in each case.  Values of  $\Phi^2$ for pentahedral clusters are also predicted by Eq.~(\ref{eqn:chidefined}), and the MSADs for these clusters can be similarly collapsed.

Knowing how uncertainty in particle positions affects measurements of dynamics quantities, such as the MSAD, is clearly important.  However, one is typically unable to define the noise level so precisely in experiments.  As stated earlier, particle tracking resolution depends on various experimental factors including particle size, optics, and the type of camera used to record images. As shown in Fig. \ref{fig:msadnoise}(a), for slowly diffusing clusters and moderate-to-high noise levels, the measurements of the MSAD may not be representative of the true dynamics until fairly large lag times.  At these large $\Delta t$, statistics are poorer than small $\Delta t$, and confident measurements are harder to obtain.  Therefore, this type of oversampling can be detrimental to microscopy experiments where data sets consist of a few hundred to a few thousand images.  Thus it is important to determine an imaging rate
that yields the most information from a data set.

In microscopy experiments, one can effectively increase the rotational signal-to-noise ratio between each time step by imaging less frequently.  Determining the appropriate sampling rates in experiments can be done by estimating the noise level $\sigma_x$, the diffusion constant $D_R$ and calculating an approximate $\Phi^2$.  Diffusive motion will begin exceeding noise when $\Phi^2 \approx 4D_R \Delta t$. Solving for $\Delta t$ then yields a reasonable sampling lag time. It is also important to avoid undersampling, that is, too long a lag time.  Doing so will make diffusive motion between images appear erroneously slow.  We find that a good rule of thumb for the upper limit on sampling time should be the time when the cluster has diffused $\approx $1 radian$^2$.  Thus, an estimate for the upper limit is $\approx 1/4D_R$.  However, as in all particle tracking experiments, the time between images must be small enough that individual particles can be confidently identified.  This typically means that particles must be imaged before moving a distance of one interparticle spacing  \cite{crocker1996jcis}.  

%
%

\section{Experimental Application}
\label{sec:experiment}

Measurements of rotational motion of diffusive tetrahedral clusters have been performed using the described method \cite{elsesserlang2011}. Real fluorescent tetrahedral clusters are synthesized as in \cite{elsesserlang2011}. A cluster is composed of individual poly(methyl methacrylate) (PMMA) spheres, each with a diameter 2.45~$\mu$m as measured by static light scattering (SLS).  The particles within a cluster are irreversibly bound together, but are sterically stabilized to prevent the possibility of aggregation to other clusters.  Dilute suspensions of clusters are prepared in a mixture of cyclohexyl bromide (CXB) and {\it cis}-decalin (DCL)  at a ratio of 85/15 (w/w) that closely matches both the density and index of refraction of the particles. Clusters are imaged in 3D over time with a Leica TCS SP5 confocal microscope.  We track locations of the individual particles within a tetrahedron using standard particle tracking routines \cite{crocker1996jcis,weeksmicroscopy1}.  The uncertainties in particle position for these experiments are $\approx 30$~nm in the $x$- and $y$-directions, and $\approx 40$~nm in the $z$-direction.  Given these tracking resolutions and assuming a maximally packed tetrahedron, from Eq. (\ref{eqn:chidefined}) we estimate the angular resolution in this experiment as $\Phi \approx 0.028$~radians ($1.6^{\circ}$).

Once tracked, we calculate the translational MSD and the MSAD and determine the translational and rotational diffusion coefficients, $D_T$ and $D_R$, respectively.   In three dimensions, the translational MSD is described by the Stokes-Einstein-Sutherland equation

\begin{equation}
\langle \Delta r^2 \rangle = 6 D_T \Delta t,
\label{eqn:ses}
\end{equation}
while the MSAD is described by Eq. (\ref{eqn:sed}).  

Hydrodynamically, tetrahedral clusters can be accurately modeled as spheres~\cite{hoffmann2009} given the relation

\begin{equation}
d_{\rm tetra} = 1.844 \times d_{\rm sphere},
\label{eqn:tetsphere}
\end{equation}
where $d_{\rm tetra}$ is the effective hydrodynamic diameter of the cluster, and $d_{\rm sphere}$ is the diameter of the particles within the cluster.  Theoretical translational and rotational diffusion coefficients, $D_T$ and $D_R$ respectively, can be calculated using

\begin{equation}
D_T = \frac{k_B T} {3 \pi \eta d_{\rm tetra}}
\label{eqn:theotransd}
\end{equation}

\begin{equation}
D_R = \frac{k_{B} T}{\pi \eta d_{\rm tetra}^3},
\label{eqn:theorotd}
\end{equation}
where $k_B$ is Boltzmann's constant, $T$ is the temperature, and $\eta$ is the viscosity of the suspending solvent.   The viscosity of the CXB/DCL mixture was measured at $\eta = 2.18$ mPas and experiments were performed at $T = 295$ K. 
\begin{figure}
\begin{center}\includegraphics[width=.9\textwidth]{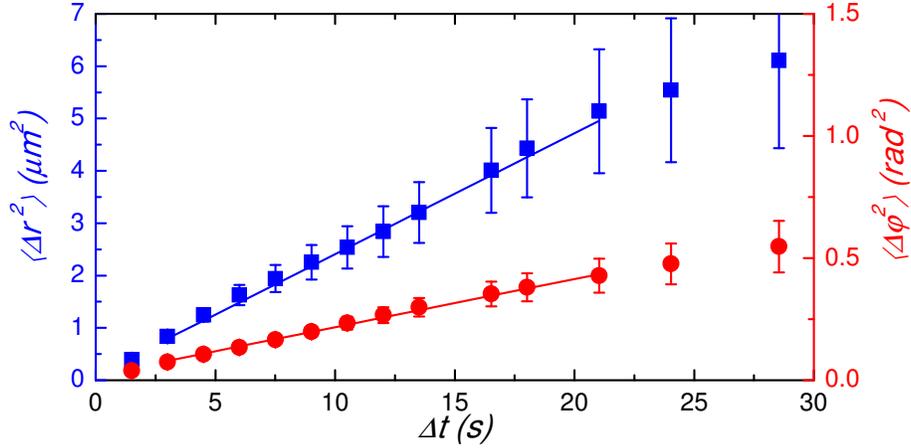}
\end{center}
\caption{Calculated MSD (blue squares) and MSAD
 (red circles) of a diffusing tetrahedral cluster.  Solid lines are best fit lines over range of data used to determine the respective diffusion coefficients.}
\label{fig:expmsds}
\end{figure}

In Fig.~\ref{fig:expmsds}, we show the MSD and MSAD of a diffusing tetradedral cluster.  The MSD corresponds to motions of the center of mass of the cluster, and the MSAD is an average over the individual particle MSADs (i.e., using the initial orientations of the particles as four separate $\hat{\mathbf{p}}^0$).  Solid lines are fits to the data over the indicated range $3~\rm{s} \leq \Delta t \leq 22~\rm{s}$ .  As can be seen, both the MSD and MSAD are approximately linear only at small lag times.  This is because the data set used to make these measurements consists of only a single cluster for less than 300 timesteps ($\approx$~220 s).  Diffusion coefficients are measured by performing linear fits to the MSD and MSAD shown in Fig.~\ref{fig:expmsds} over the indicated range and extracting the slopes of the lines.   

\begin{table}[h! t!]
 \caption{Measurements from Tracking Tetrahedral Cluster}
 \centering 
\begin{tabular}{l  c  c  r} 
Diffusion & Coefficient & $d_{\rm tetra}~(\mu{\rm m})$ & $d_{\rm sphere}~(\mu{\rm m})$
\\[.25pt] 
\cline{1-4}
 \\[-8pt]
Trans. & ($3.85 \pm 0.49) \times 10^{-2}~\mu\mathrm{m}^2$/s & $5.15 \pm 0.67$ & $2.79 \pm 0.13$\\
Rot. & $(4.93 \pm 0.49) \times 10^{-3}~\mathrm{rad}^2$/s & $4.94 \pm 0.20$ & $2.68 \pm 0.11$\\
\hline
\end{tabular} 
\label{tab:expmeasure} 
\end{table}

Shown in Table \ref{tab:expmeasure} are the measured translational and rotational diffusion coefficients.  Values of $d_{\rm tetra}$ are calculated from Eqs. (\ref{eqn:theotransd}) \& (\ref{eqn:theorotd}) using known experimental conditions, and values for $d_{\rm sphere}$ then follow trivially using Eq. (\ref{eqn:tetsphere}).  Ideally, the sizes calculated from translational motions would be identical to those calculated from rotational motions.   As shown, these values agree to within 4\%.  Such good agreement between these two measurements demonstrates our ability to track translational and rotational motions of clusters simultaneously.

Previous work using standard particle tracking and our matrix method reported particle diameters $\approx 14$\% larger  than the 2.45~$\mu$m measured by SLS~\cite{elsesserlang2011} .  A reanalysis of the data, presented above, improves the particle diameter measured using our matrix method to within $\approx 9$\%.  We attribute the better accuracy of using rotational motion, as opposed to translational, to the least squares minimization process, which incorporates the rotational motions of all particles within a cluster.  The remaining disagreement is likely due to multiple factors, including the fitting routines used in SLS and swelling of particles when in a solvent of CXB/DCL.  SLS measurements were performed with suspensions in pure DCL, which is not known to swell particles.

\section{Conclusion}\label{sec:conclusion}
We have presented a simple method of tracking the rotational motions of colloidal clusters.  Our method implements conventional particle tracking routines to determine the locations of individual particles within a cluster and uses this information to compute rigid body transformations that describe changes in a cluster's orientation over time.  The set of matrix transformations constitutes a global description of a cluster's motion during the course of an experiment or simulation and allows one to calculate rotational dynamics about any arbitrary axis.  The least squares minimization used in this method considers the motions of each individual particle in calculating the rotation matrix, and therefore measurements of rotational displacements are less sensitive to tracking noise.  Additionally, precision in determining angular displacements increases with cluster size $R$.  When combined, the resolution of measuring angular displacements scales with 1/$nR^2$.

Though constructed specifically for tracking colloidal clusters, we emphasize that our method is not limited to these systems, but is applicable to following the rotational motion of any body over time, providing that at least three distinct noncolinear points in the body can be reliably distinguished.    Because the accuracy of tracking a rotation depends on the number of particles and the size of the body, the possibility exists that this method can be adapted to many-body systems where each particle diffuses independently of the rest while the system itself also undergoes bulk rotations.  In this scenario, diffusive motions are random and can therefore be treated as noise while calculating $\{\mathbf{R}^k\}$ for the system. The rotation matrices can be inverted and, in a manner similar to Eq. (\ref{eqn:rotprod}), the bulk rotations can be removed, leaving only uncorrelated diffusive motion.  Thus, this method is also applicable to determining particle motions in rotating coordinate frames.

\section*{Acknowledgements}
G.L.H., K.V.E., and E.R.W. acknowledge financial support from NSF Grant CHE-0910707, and M.T.E. acknowledges support from NSF Grant DMR-0706453.  We thank D.J. Pine for helpful discussions.

\section*{Appendix}\label{sec:rotviarot}

Given a rotation matrix $\mathbf{R}$, one can calculate at each time the axis of rotation $\hat{\mathbf{u}}$ and angular displacement  $\Delta \upvarphi$.  In this notation, a rotational displacement can be described by a vector $\Delta \vec{\upvarphi} = \Delta \upvarphi \hat{\mathbf{u}}$, where $\hat{\mathbf{u}}$ has components in each of the Cartesian axes.

A single rotation will, by definition, have no effect on the direction of $\hat{\mathbf{u}}$, therefore,

\begin{equation}
\mathbf{R}\hat{\mathbf{u}} = \hat{\mathbf{u}}.
\label{eqn:Ronu}
\end{equation}

\noindent From Eq. (\ref{eqn:Ronu}), we see that the axis of rotation is an eigenvector of the matrix $\mathbf{R}$ with an eigenvalue of 1.  For a set of rotational displacements $\{\mathbf{R}^k\}$, one may determine the axes of rotation by calculating the eigenvectors and eigenvalues of the rotation matrices, searching for the eigenvalues equal to 1, and taking the corresponding eigenvectors.

To determine the size of the displacement about the axis of rotation, one defines an arbitrary vector $\hat{\mathbf{w}}$ perpendicular to $\hat{\mathbf{u}}$.  For simplicity, we choose $\hat{\mathbf{w}}$ to be perpendicular to the $x$-axis (denoted by $\hat{\mathbf{i}}$),

\begin{equation}
\label{eqn:defv}
\hat{\mathbf{w}} = \frac{\hat{\mathbf{u}} \times \hat{\mathbf{i}}}{ |\hat{\mathbf{u}} \times \hat{\mathbf{i}} | }.
\end{equation}
and apply the rotation matrix,
\begin{equation}
\mathbf{R}\hat{\mathbf{w}} = \hat{\mathbf{w}}'.
\label{eqn:Ronv}
\end{equation}

\noindent The magnitude of the displacement is the angle between $\hat{\mathbf{w}}$ and $\hat{\mathbf{w}}'$, and can be computed using the cross product relation

\begin{equation}
\sin{\left( \Delta \upvarphi \right) } \hat{\mathbf{u}} =  \hat{\mathbf{w}}  \times \hat{\mathbf{w}}'.
\end{equation}

Calculating displacements relative to the axis of rotation always results in displacements greater than or equal to those measured relative to an arbitrary $\hat{\mathbf{p}}^0$.  For example, measuring a diffusion coefficient of a spherically symmetric body with this method will yield a value that is a factor of $3/2$ of the actual diffusion coefficient in Eq. (\ref{eqn:sed}).

\begin{figure}[h]
\begin{center}\includegraphics[width=.99\textwidth]{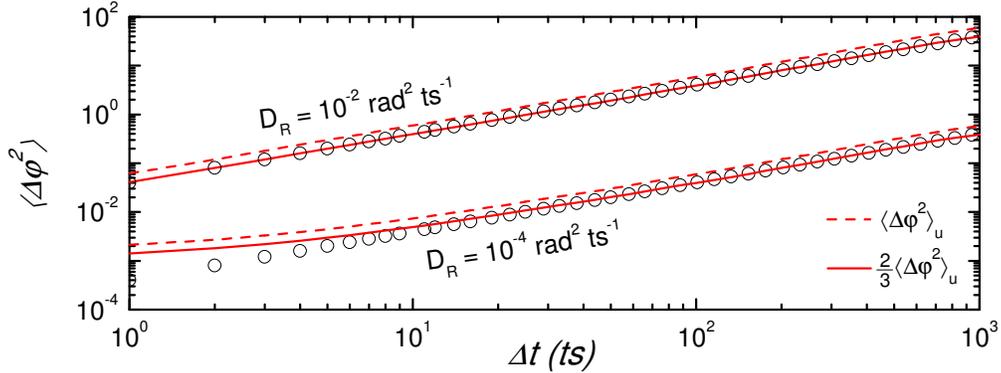}
\end{center}
\caption{The MSADs of pentahedra with diffusion coefficients $D_R = 10^{-4}$~rad$^2/$ts and $D_R = 10^{-2}$~rad$^2/$ts and $\sigma_x = $ 50 nm.  Dashed lines show the uncorrected MSAD described below and solid lines are corrected data, obtained by multiplying the dashed lines by 2/3. Open circles are the theoretical MSAD from Eq. (\ref{eqn:sed}).}
\label{fig:pmsad}
\end{figure}

To understand the origin of this difference, consider the rotational axis $\hat{\mathbf{u}} = \hat{\mathbf{z}}$, a perpendicular vector $\hat{\mathbf{w}} = \hat{\mathbf{x}}$, and an arbitrary vector $\hat{\mathbf{p}}^0$ located at spherical coordinates $(\sin{\theta},0,\cos{\theta})$.  If $\hat{\mathbf{w}}$ is rotated by an amount $\delta \upvarphi_u$, the angle between  $\hat{\mathbf{w}}$ and  $\hat{\mathbf{w}}'$ is identically $\delta \upvarphi_u$.  However, the angle $\delta \upvarphi$  between $\hat{\mathbf{p}}$ and $\hat{\mathbf{p}}'$ can be shown to be

\begin{equation}
\label{eqn:angdiff}
\delta \upvarphi = \cos^{-1}{\left( \sin^2{\theta}\cos{\delta \upvarphi_u} + \cos^2{\theta} \right)}
\end{equation}

For small $\delta \upvarphi_u$, we can approximate $\delta \upvarphi$ as

\begin{equation}
\label{eqn:angapprox}
1-\delta \upvarphi^2 \approx \sin^2{\theta} \left( 1-\delta \upvarphi_u^2 \right) + \left( 1 - \sin^2{\theta} \right)
\end{equation}

\begin{equation}
\label{eqn:angapprox2}
\delta \upvarphi^2 \approx \sin^2{\theta}\delta \upvarphi_u^2.
\end{equation}

If an average of Eq. (\ref{eqn:angapprox2}) is taken over spherical coordinates, we are left with an expression similar to an MSAD, 
\begin{equation}
\label{eqn:angapprox3}
\langle \delta \upvarphi^2 \rangle =  \langle \sin^2{\theta} \rangle \langle \delta \upvarphi_u^2 \rangle = \frac{2}{3} \langle \delta \upvarphi_u^2 \rangle,
\end{equation}

The factor of $2/3$ arises for the same reasons in Perrin's original derivation of rotational diffusion~\cite{perrin1928}.  Fig. \ref{fig:pmsad} shows the MSADs measured in this way for two pentahedral clusters.  As stated, one can correct for the overestimation of motion by simply multiplying the MSAD by 2/3.  The average MSAD for any body can be corrected in the same manner, however we stress that this method returns only the average dynamics.  While more direct than using an orientation vector, characterizing anisotropic bodies in this way will convolute motions about separate axes, and so will require some care when interpreting results.\\

\end{document}